\documentclass{article}

\usepackage{amssymb}
\usepackage{epsfig,psfrag}
\usepackage{dcolumn}
\usepackage{bm}
\usepackage{epsfig} 
\usepackage{graphicx}
\usepackage{epstopdf}
\usepackage{color}
\usepackage{array}
\usepackage{siunitx}

\usepackage[hmargin=3.5cm,vmargin=3.5cm]{geometry}
\begin{document}
\title{Terahertz emission from multiple-microcavity exciton-polariton lasers}
\author{ S. Huppert, O. Lafont, E. Baudin, J. Tignon and R. Ferreira \\
\textit{Laboratoire Pierre Aigrain, Ecole Normale Sup\'erieure-PSL Research University, }\\
\textit{CNRS, Universit\'e Pierre et Marie Curie-Sorbonne Universit\'es } \\
\textit{ Universit\'e Paris Diderot-Sorbonne Paris Cit\'e, } \\
\textit{24 rue Lhomond, 75231 Paris Cedex 05, France}}

\date{}
\maketitle
\begin{abstract}
Terahertz emission between exciton-polariton branches in semiconductor microcavities is expected to be strongly stimulated in the 
polariton laser regime, due to the high density of particles in the lower state (final state stimulation effect). 
However, non-radiative scattering processes depopulate the upper state and greatly hinder the efficiency of such terahertz sources. 
In this work, we suggest a new scheme using multiple microcavities and exploiting the transition between two interband polariton branches located below the exciton level. 
We compare the non-radiative processes loss rates in single and double cavity devices and we show that a dramatic reduction can be achieved in the latter, 
enhancing the efficiency of the terahertz emission. 
\end{abstract}

The development of efficient coherent sources for terahertz (THz) emission is a subject of intense activity for both applied and fundamental research. Various strategies have been 
exploited, in particular using nonlinear optical effects such as frequency multiplication in nonlinear diodes \cite{pearson2011THz_GHz_multiplication} or difference frequency generation 
from infrared or optical laser sources \cite{auston1984photoconducting_antenna}. A strong effort has also been devoted to extending the use of quantum cascade laser to the 
THz range \cite{kohler2002terahertz_QCL} but the realization of powerful and compact sources in that frequency domain remains challenging. 
Recently, it was suggested that exciton-polariton lasers could be an efficient tool for THz generation \cite{Kavokin_champE, Kavokin_sp, Ciuti_PRB2013_AQW}. 
The strong coupling between quantum well excitons and photons in a semiconductor microcavity produces mixed light-matter states called 
exciton-polaritons \cite{weisbuch1992prl_polariton}. Under intense optical excitation, the polariton lasing regime is reached and the polariton population of the lower branch 
becomes extremely high in the vicinity of the zero momentum state \cite{Imamoglu1996theory_polariton_laser, kasprzak2006BEC_polariton}. 
Radiative transitions between the upper and the lower polariton branches are strongly favored in this regime due to final state stimulation effect. Furthermore, the Rabi splitting 
between the two branches
is typically several meV, which makes polariton lasers very promising for THz generation (1 THz corresponds to 4.1 meV). 
However, for centrosymmetric quantum well structures, the dipole matrix element between upper and lower polaritons is zero and the THz transition is symmetry-forbidden. 
In order to allow for THz emission, it was suggested to break the centrosymmetry by applying an electric field \cite{Kavokin_champE}, to populate the $1p$-exciton state with 
two-photon pumping \cite{Kavokin_sp} or to realize structures in which $1s$ and $2p$ excitons are resonant and hybridize \cite{Kavokin_PRL2013_Bosonic_cascade}. Until now, 
these three approaches have not brought any experimental confirmation of THz emission. 

\begin{figure}[!ht]
\centering
\includegraphics[scale=1]{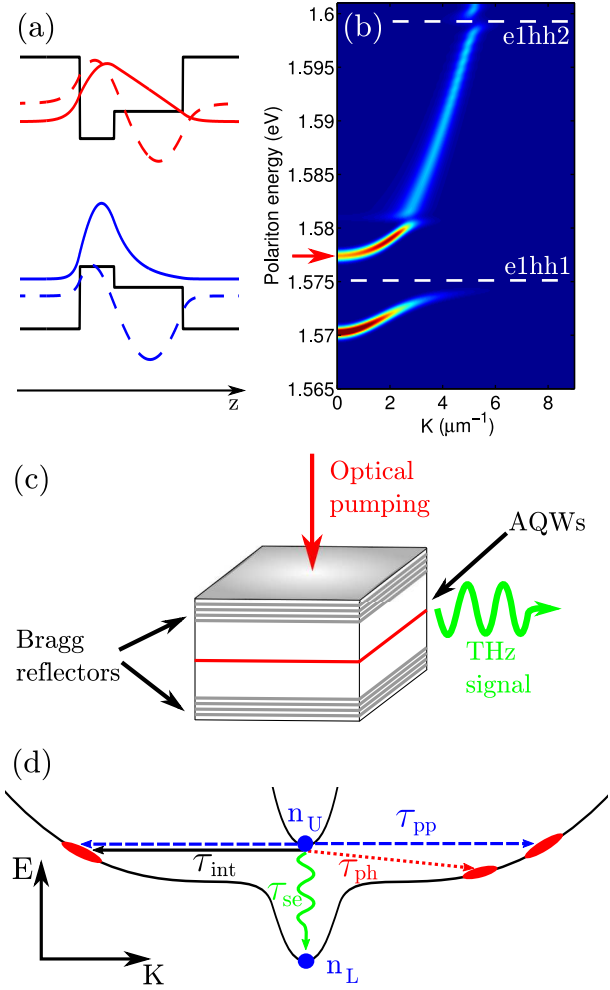}
\caption{(a) Band structure of the AQW along the growth axis $z$, together with the electron (red) and heavy hole (blue) wavefunctions of the states 1 (full line) and 
2 (dashed line). 
(b) Computed dispersion spectrum for 9 identical AQWs placed in a single microcavity. The color scale shows the photonic part of the polariton states. The 
white dashed lines indicate the energy of the bare $e1hh1$ and $e1hh2$ excitons and the red arrow points to the optically pumped state.
(c) Schematic representation of the microcavity with the AQWs layer (red line), also showing the pump and THz beams.
(d) Schematic polariton band structure for the single microcavity scheme. The THz transition is represented by 
a green vertical arrow, scattering by interface roughness and acoustic phonons are indicated with full black and dotted red arrows respectively. 
Horizontal polariton-polariton collisions are represented by the two dashed blue arrows.}
\label{Fig1cav}
\end{figure}

The present work focuses on using a quantum well 
with intrinsically asymmetric design and shows that the THz transition becomes allowed in such a structure \cite{Ciuti_PRB2013_AQW}.
This configuration provides efficient optical pumping of the upper state and non-zero THz emission 
probability to the lower state, without complexifying the experimental set-up. 
The main obstacle to the THz emission is the existence of non-radiative scattering processes which deplete the upper polariton state 
very efficiently (see figure \ref{Fig1cav}.d). Following Diederichs \textit{et al.} \cite{diederichs2005multicav, Tignon_multiple}, we suggest a new device using multiple microcavities 
in order to hinder non-radiative scattering. We compare the scattering rates associated with three different mechanisms in single and in double microcavities, and we show that 
these rates can be reduced by four orders of magnitude in the latter configuration, while the THz emission rate is only weakly affected. In that case,  
the device efficiency is no longer limited by non-radiative scattering but only by the radiative decay of injected polaritons. The conversion ratio of polaritons into 
THz photons is predicted to reach $10^{-3}$, which is promising for THz technologies, as room temperature polariton lasing was demonstrated in several experimental studies 
\cite{christopoulos2007roomT_PL, christmann2008RT_PL_QW, Bhattacharya2014electrically_pumped_polariton_laser_RT}.

Figure \ref{Fig1cav} shows the single cavity scheme that is considered: 
the quantum wells are placed inside an optical microcavity, and pumped optically through one of the Bragg reflectors (our THz generation scheme is also compatible with electrically 
pumped polariton lasers \cite{schneider2013electrically_pumped_polariton_laser, Bhattacharya2014electrically_pumped_polariton_laser_RT}). 
For centrosymmetric quantum well structures, direct radiative transitions between upper and lower polariton branches are forbidden  
because the bright excitons are those involving electron and hole states of the same parity (e.g. $e1hh1$, 
$e2hh2$ or $e1hh3$ where $e$n and $hh$m label the $n$-th electron and the $m$-th heavy hole quantum well states), and on the contrary, THz transitions change the parity of one of the 
particles, while conserving the state of the other. Thus THz transitions between bright polaritons are symmetry-forbidden.
Therefore in this work we focus on asymmetric quantum well (AQW) structures in which these selection rules are suppressed. 
A wide range of AQW parameters were tested in order to provide sizable THz emission. The calculations presented in the following were performed for   
structures composed of a 5 nm GaAs layer and a 10 nm Al$_{0.05}$Ga$_{0.95}$As layer embedded in Al$_{0.3}$Ga$_{0.7}$As barriers. 
Figure \ref{Fig1cav}.a. shows the schematic band structure of these AQWs as well as the wavefunctions of the two lowest electron and heavy hole states. 
In this non-centrosymmetric structure, odd excitons $e1hh2$ and $e2hh1$ are bright 
and introduce a sizable dipole for THz emission. The dipole is oriented along the growth axis, therefore the radiation is polarized along this axis 
and emitted mainly from the sides of the sample.

\begin{figure}[!ht]
\centering
\includegraphics[scale=1]{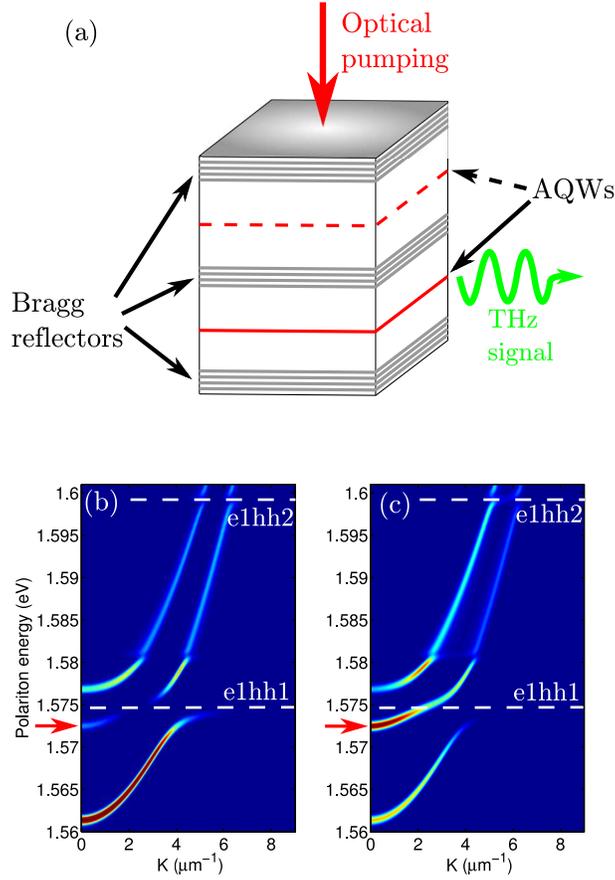}
\caption{(a) Schematic representation of the double microcavity with the AQWs layers (red lines). In order to generate THz radiation, the AQWs layer of the top cavity should be removed 
(dashed red line). (b,c) Computed dispersion spectrum for 9 identical AQWs placed in the bottom microcavity. 
The color scale shows the photonic part of the polariton states inside the bottom (b) and top cavity (c). The 
white dashed lines indicate the energy of the bare $e1hh1$ and $e1hh2$ excitons and the red arrow points to the optically pumped state. 
The energy splitting between the two bare cavity modes (without AQW) is $12$ meV.}
\label{Fig2cav}
\end{figure}

To compute the eigenstates of the AQW in the strong coupling regime and calculate the polaritonic dispersions presented in fig.\ref{Fig1cav} and \ref{Fig2cav}, 
the hamiltonian of the system is diagonalized numerically for each value of $\mathbf K$ on a truncated basis consisting of 
the cavity mode and of the electron-hole states $|n_e, n_h , m, \mathbf K \rangle$ whose two-particles wave-function is defined as: 
$$
\Psi_{n_e, n_h , m, \mathbf K} (z_e,z_h, \mathbf R, \boldsymbol \rho) =\frac{e^{i \mathbf K. \mathbf R}}{\sqrt S} \ \phi_{n_e}(z_e) \  \phi_{n_h}(z_h) \  \psi_{m}(\rho),
$$
where $\phi_{n_e}$ and $\phi_{n_h}$ are the wavefunctions of the electron and hole states along the growth axis (shown on Fig. \ref{Fig1cav}.a), $\mathbf R$ and 
$\boldsymbol \rho$ are the center-of-mass and relative in-plane coordinates and $S$ is the sample area. The functions $\psi_m$ are defined on a circular confining box of radius 
$R_0$:
$$
\psi_{m}(\rho) = N_{m} J_0(k_{m} \rho),
$$
with $N_{m}$ a normalization factor and $k_{m}$ such that $J_0(k_{m} R_0)=0$. 
Only s-symmetry excitons are included, as non-zero angular momentum states do not couple to light. 
The hamiltonian includes kinetic contribution, quantum well potential, electron-hole Coulomb interaction as well as light-matter coupling. The coupling between the cavity mode 
and the state $|n_e, n_h , m, \mathbf K \rangle$ is expressed in the enveloppe function approximation as $\gamma \langle \phi_{n_e} | \phi_{n_h} \rangle \psi_{m}(\rho=0)$, where the 
coupling strength $\gamma$ is a constant parameter proportional to the electric field of the cavity mode at the location of the quantum wells. This field depends on the cavity design and 
in this study, the parameter $\gamma$ was fixed according to typical experimental values for GaAs/AlGaAs structures.  

In the non-centrosymmetric quantum well of figure \ref{Fig1cav}.a, there is no selection rule on $n_e$ and $n_h$ and all electron-hole transitions are bright. 
The states up to $n_e=2$ and $n_h=2$ are included and both $e1hh2$ and $e2hh1$ excitons yield comparable contributions to the THz dipole. Furthermore, we checked that 
the higher states do not contribute significantly to the THz emission.   
Figure \ref{Fig1cav}.b shows the polariton dispersion calculated for the AQW of Fig. \ref{Fig1cav}.a. The color plot shows the photonic part of the polaritons (which 
can be accessed experimentally through transmission or reflection measurements). At low wavevector $K$, the cavity is close to resonance with the $e1hh1$ exciton 
and two main polariton branches are observed, separated by the Rabi splitting.  
Above 1.58 eV, the cavity mode is broadened due to the weak coupling with the $e1hh1$ continuum and it anticrosses with the $e1hh2$ exciton at energy 1.6 eV.

The spontaneous emission rate of THz photons between the two polariton states $|U \rangle$ and $|L \rangle$ (see Fig. \ref{Fig1cav}.d) is given by:
\begin{eqnarray*}
R_{U\rightarrow L} = n_U (n_L+1) W =\frac{n_U}{\tau_{\text{se}}}, 
\end{eqnarray*}
with $n_U$, $n_L$ the population in $|U \rangle$ and $|L \rangle$ states respectively and, 
$$
W=\frac{\omega^3 e^2 n |\langle U |\hat z|L \rangle |^2}{3 \pi \epsilon_0 \hslash c^3},
$$
where $\omega$ is the pulsation of the THz radiation and $n$ is the refractive index of GaAs in the considered frequency range.
Interestingly, due to the bosonic nature of polaritons, the radiated THz power is proportional to the occupation of the lower mode $n_L+1$. Thus in the polariton lasing regime 
\cite{bajoni2008Bloch_micropillar_polariton, kasprzak2006BEC_polariton}, 
we expect a strong enhancement of the spontaneous emission through final state stimulation effect, as the modes near $\mathbf K=0$ become macroscopically occupied, up to $n_L=10^4$. 
For this occupancy, the emission time $\tau_{\text{se}}$ is reduced below $\SI{1}{\micro s}$. 
It can be furthermore reduced by two orders of magnitude due to Purcell effect if the device is placed in a THz cavity \cite{Kavokin_champE}. In that case, 
as the interaction time with the THz photons is increased, reabsorption from the lower state and stimulated emission should also be taken into account. 
The net effect of these two processes is favorable to the emission only if $n_L<n_U$, but this population inversion is difficult to 
achieve for polaritons, as the non-radiative relaxation processes are very efficient (see discussion below). For the sake of simplicity, in the following all calculations are 
performed assuming that the THz field is not confined.

As previously mentioned, the emission of THz photons competes with efficient non-radiative loss mechanisms. Polariton scattering processes have been widely investigated 
\cite{Tassone_PRB1999_Polariton_interaction, Ciuti_PRB2000_Pol_pol_int,Tassone_PRB1996_Polariton_phonon}, 
and the associated rates can be readily obtained from the polaritonic states determined numerically.
In this work, we focus on the scattering of polaritons off 
the upper state arising from three processes illustrated on Fig. \ref{Fig1cav}.d: acoustic phonons, interface roughness and horizontal polariton-polariton 
scattering. The polariton-polariton scattering rate was calculated in the 
approximation of long wavelength of the final state $K_f^{-1}\gg a_B$ with $a_B$ the Bohr radius of the $1s$ exciton \cite{Tassone_PRB1999_Polariton_interaction, Ciuti_PRB2000_Pol_pol_int}. 
It is the dominant scattering mechanism, with sub-picosecond characteristic time $\tau_{\text{pp}}$ at density $N_L=10^9\ $cm$^{-2}$ per quantum well (population in the upper band in the 
vicinity of $\mathbf K=0$), which corresponds to the typical order of value obtained at the polariton lasing threshold \cite{bajoni2008Bloch_micropillar_polariton, kasprzak2006BEC_polariton}. 
Acoustic phonons scattering time $\tau_{\text{ph}}$, calculated in the standard approach \cite{Tassone_PRB1996_Polariton_phonon}, is found to be about 250 ps at low temperature. 
Interface roughness scattering was also evaluated generalizing a model previously developed for 
quantum cascade structures \cite{ndebeka2014revue}. This process was found to scatter surprisingly efficiently, 
with a characteristic time $\tau_{\text{int}}\simeq 30\ $ps for 1\% covering of the interfaces. The efficiency of this last mechanism is mainly due to the presence of the step in 
the AQW potential (see Fig. \ref{Fig1cav}.a)  which introduces roughness at a place where the \textit{e1}
and \textit{hh1} states have a high probability of presence. 

The estimations above show that scattering seriously hinders the THz emission in the single cavity scheme, mainly through 
polariton-polariton interactions. However, it has been shown theoretically and experimentally \cite{diederichs2005multicav, Tignon_multiple} that a significant reduction of 
the losses can be achieved by the use of multiple microcavities. 
Indeed, all three scattering rates mentioned above are very sensitive to the density of available final states. For a single cavity near resonance, this density is fixed 
by the exciton dispersion and is roughly independent from the detuning. In return, with a negatively detuned multiple cavity, 
both states $|U\rangle $ and $|L\rangle$ can be situated at lower energy than the excitonic reservoir. In this case, the 
density of final states is given by the steep polariton dispersion and as we show below, the scattering is dramatically suppressed.

In the following, we discuss the double cavity scheme represented in Fig. \ref{Fig2cav}.a. The most natural choice would be to put identical AQWs in the two cavities and in 
the same number.
\begin{figure}[!ht]
\centering
\includegraphics[scale=1]{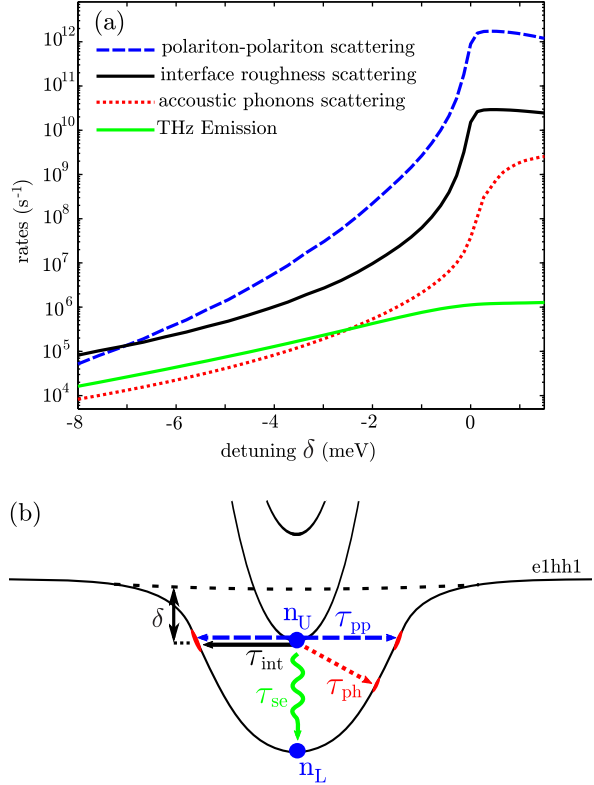}
\caption{(a) THz emission (green), low temperature acoustic phonon scattering (red) interface roughness scattering (black) and polariton-polariton collision rates (blue) are 
plotted as a function of the detuning $\delta$ between the polariton state $|U\rangle$ and the excitonic reservoir $e1hh1.$
(b) Schematic representation of the polariton dispersion for the double cavity of Fig. \ref{Fig2cav} at negative $\delta$. The considered scattering processes are represented 
as in Fig.\ref{Fig1cav}.d. The density of final states after scattering is strongly reduced compared to the single cavity scheme.}
\label{scattering}
\end{figure}
However in that case, 
the polariton states are either symmetric or antisymmetric for both their photonic and their excitonic parts. This leads to the appearance of a new selection rule 
for the inter-polariton transitions which forbids THz emission between the two levels situated below the excitonic reservoir, because they have opposite symmetry. 
This selection rule holds, for any number of cavities and for any coupling and any detuning between the photon modes, assuming that identical AQWs are placed in all the cavities. 
In order to suppress this detrimental selection rule, we suggest to place AQWs in one of the two cavities only (the wells indicated with a dashed line in Fig. \ref{Fig2cav}.a are removed). 
The resulting dispersions are shown on Fig. \ref{Fig2cav}.b and \ref{Fig2cav}.c, considering the same AQWs as in Fig. \ref{Fig1cav}.
The upper photonic branch is resonant with the 
$e1hh1$ exciton and the Rabi splitting is reduced by a factor 2 with respect to the one in Fig. \ref{Fig1cav}, as the AQWs are placed only in one of the cavities. 
It is also noticeable that the AQWs should be placed in the bottom microcavity, as sketched on Fig. \ref{Fig2cav}.a, in order for the upper polariton 
$|U\rangle $ to be efficiently pumped. 
Indeed, state $|U\rangle $ has a high probability of presence in the top cavity but only a low one in the other cavity, which contains the wells. The reverse 
is true for $|L\rangle $, which should therefore have a long radiative lifetime and easily reach the polariton-lasing regime.

In the double cavity scheme, the scattering rates depend dramatically on the detuning $\delta$ (defined on Fig. \ref{scattering}.b) between initial state $|U\rangle$ and $e1hh1$ exciton, 
as shown in Fig. \ref{scattering}.a. Indeed if the polariton $|U\rangle$ 
is above the excitonic reservoir ($\delta>0$), the THz emission rate and the scattering rates are very similar to that obtained in the single cavity scheme. 
However, when $\delta$ becomes negative, the scattering rates decrease abruptly while the THz emission remains essentially unchanged. 
This is because the density of final states available for the scattered polaritons strongly 
decreases when $|U\rangle$ falls below the bare exciton energy. Further decrease of $\delta$ slowly reduces all four rates because the excitonic part of polaritons 
$|U\rangle$ and $|L\rangle$ decreases. In order to achieve efficient THz generation, an intermediate detuning should be chosen in order to suppress significantly the non-radiative 
losses while maintaining a high THz emission rate. For instance, for $\delta=-3\ $meV, as in Fig. \ref{Fig2cav}.b and \ref{Fig2cav}.c, the losses by scattering are reduced by 
four orders of magnitude with respect to the single cavity case while the THz emission rates is reduced by less than a factor 10. 
Furthermore, the ratio between THz emission and polariton-polariton scattering reaches 1\% at $\delta=-3\ $meV. 
In that case, the non-radiative scattering rates become negligible compared to the radiative decay rate, which is typically of the order of $10^{11}$ s$^{-1}$. 
Therefore radiative escape of the polaritons out of the cavity becomes the dominant loss mechanism, and the ratio of conversion of the injected polaritons into THz photons reaches $10^{-3}$ 
if the system is placed in a resonant THz cavity (assuming Purcell factor of 100).

In conclusion, we have compared THz emission from AQWs in single and in multiple microcavities. In the latter case we have derived a new selection rule imposing that the content of all 
cavities should not be identical. We have evaluated the losses arising from three different non-radiative processes and shown that in a double cavity under appropriate conditions, 
the scattering rates can be reduced by four orders of magnitude with respect to single cavity devices.

\bibliographystyle{unsrt}

\end{document}